# Network Coding Channel Virtualization Schemes for Satellite Multicast Communications

Samah A. M. Ghanem[†], Ala Eddine Gharsellaoui[*], Daniele Tarchi[*] and Alessandro Vanelli-Coralli[*]
[*]Department of Electrical and Electronic Engineering, University of Bologna, Italy
[†]Independent Senior Researcher

*Abstract*—In this paper, we propose two novel schemes to solve the problem of finding a quasi-optimal number of coded packets to multicast to a set of independent wireless receivers suffering different channel conditions. In particular, we propose two network channel virtualization schemes that allow for representing the set of intended receivers in a multicast group to be virtualized as one receiver. Such approach allows for a transmission scheme not only adapted to per-receiver channel variation over time, but to the network-virtualized channel representing all receivers in the multicast group. The first scheme capitalizes on a maximum erasure criterion introduced via the creation of a virtual worst per receiver per slot reference channel of the network. The second scheme capitalizes on a maximum completion time criterion by the use of the worst performing receiver channel as a virtual reference to the network. We apply such schemes to a GEO satellite scenario. We demonstrate the benefits of the proposed schemes comparing them to a per-receiver point-to-point adaptive strategy.

## I. INTRODUCTION

Multicast communications are fundamental to many practical applications, including Satellite TV broadcast, content delivery and interactive communications, multimedia conferencing, across wired or wireless medium. Network coding is a key enabling technology that offers a unique technique to multicast communications. In particular, by mixing the information content shared among receivers, higher reliability and less delay can be encountered.

In wired networks, network coding was shown to achieve the multicast capacity [1]. In [2], it is shown that an explicit construction of a code that achieves multicast network capacity is a linear network code. In [3], Random Linear Network Coding (RLNC) was proposed for multicast, a distributed network coding approach with nodes independently and randomly selecting linear coding coefficients from inputs onto output links over a Finite Field of known size, which achieves capacity with very high probability.

In wireless networks, owing to their broadcast nature, multicasting on wireless links, suffering different channel behavior, noise, and interference levels, becomes a challenge. Since there are no explicit models that express a wireless network capacity, this is considered as an open problem; thus, the characterization of optimal approaches, that jointly minimize the system completion time to several entities, is yet an open problem. This paper goes in this direction by finding the optimal number of coded packets to transmit to all receivers of a multicast wireless network group.

Since then, network coding has shown many benefits in wireless mesh networks. Some practical network coding schemes include COPE, a XOR-form [4], or MORE, an RLNC-form [5], of random mixing of packets, or a combination of both forms [6]. Additionally, in P2P networks, network coding shows significant benefits in content distribution [7] and streaming [8]. In [9] multicast network coding capacity was shown to be inversely proportional to the connection probability among the receiving nodes in the multicast. In [10], the authors characterize the expected number of transmissions per packet and quantify its gain with network coding analytically. They have also conjecture that network coding achieves a logarithmic gain in the expected number of transmissions/retransmissions to multicast compared to an ARQ scheme.

For line networks [11], the author shows gains of RLNC and adaptive RLNC schemes compared to ARQ for time variant channels. Then the authors in [12], [13] show that adaptive RLNC in line networks can achieve different energy and rate gains adapting to the channel variation. However, those works did not consider how to jointly optimize the transmission to a set of multicast group with different channel variations per receiver.

To the best of our knowledge, this work is the first to propose schemes that can jointly design the number of coded packets to multicast to a set of multicast group receivers encountering different time variant channels.

In this paper, we try to provide in an innovative way solutions to the question *how can we optimize jointly the coded transmission to a set of receivers in a multicast group?* In particular, we try to solve such an open problem due to lack of available models that express a correlated structure, by looking at the multicast approach through a virtual network that expresses an equivalent network of a min-cut-like time variant capacity. This network virtual link represents the multicast group time variant channel and can be exploited to design optimal or near optimal number of coded packets to transmit to all receivers in a multicast group.

We capitalize on the model in [11] for coded packet transmissions over time variant channels in a line network to provide an approach for the network coding multicast problem that can express multiple receivers in the multicast group separately as point-to-point time varying channels. In particular, we propose two schemes for network coding wireless multicast, one that creates a virtual worst channel that intersects with worst case receiver channels, and another that assigns the virtual worst channel of the receiver having maximum completion time. We focus on the GEO satellite application, and consider to multicast a shared content, generated via RLNC, to a set of receivers.

## II. SYSTEM MODEL

Consider a downlink multicast over a wireless channel. The GEO satellite, as a source, performs RLNC [3]. The coefficients used to generate coded packets are chosen at random from a Galois finite field of very large size. With this, the probability of generating linearly dependent packets decreases with the field size increase. Therefore, we assume a very large field size that allows with almost probability 1 the decorrelation in the generation. However, it is worth to note that there yet exists a dependency in the probability of packet erasure due to channel variation over time which inherently exists in the Land Mobile Satellite (LMS) channel model herein considered [14]. Additionally, the erasure probability of acknowledgement packets is considered to be zero for simplicity. After RLNC, the GEO satellite multicasts to a group of $K$ receivers that should receive common content. Therefore, the per-receiver $k$ will receive a signal modeled as:

$$y_k(t) = h_k(t) \cdot A \cdot x(t) + n_k(t), \quad (1)$$

where $x(t)$ and $y_k(t)$ correspond to the transmit and receive symbols, respectively. By assuming a generalized fading model, where fading is time varying and follows the LMS model in a low height building environment [14], $A$ is the transmitted signal amplitude from the GEO satellite, $n(t)$ is the zero mean complex white Gaussian noise. While $t$ is considered to be within interval $\mathcal{T} = [0, \tau]$. Moreover, a set of $K$ receivers $\mathcal{K} = \{1, \ldots, K\}$, in the group of multicast, are associated to a vector of channel gains of length $\tau$. We assume that there is no channel coding within the received packets. Thus, for each packet, every symbol needs to be received. Therefore, the corresponding packet erasure probability of channel gain $h_k(t)$ is expressed at time instant $t \in \mathcal{T}$ as:

$$P_e(h_k(t)) = 1 - (1 - P_b(h_k(t)))^B,$$

where $P_b(h_k(t))$ is the bit error probability for a given modulation scheme, considering channel gain $h_k(t)$, and $B$ is the number of bits per coded packet.

## III. CHANNEL VIRTUALIZATION SCHEMES

We propose a novel channel virtualization approach to address the multicast modeling problem with time variant channels. Such virtualization inherently exploits the necessity to represent the wireless network in an equivalent form. Such an equivalent form allows to characterize the capacity of each wireless network by its min-cut [15]. A min-cut mimics the maximum delay or maximum erasures that limit the information flow in a wireless network. Therefore, the solution that can be proposed for a virtualized network should allow receivers with worst channel conditions to yet be able to decode received information in a reliable way.

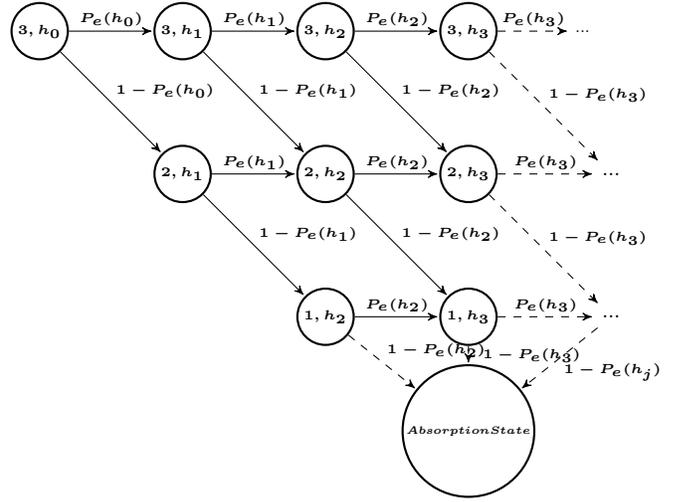

Fig. 1. Time Varying Channel Model of 3 Packets Transmission in [11]

Due to the broadcast nature of wireless communications, and the impairments associated to challenging wireless environments, like the satellite communications addressed here, the interpretation of such approach for time variant channels is to represent all the wireless links as point-to-point links with a reference virtual channel that provides most possible losses in terms of packet erasures or in terms of completion time, where both are associated to resource wastage in retransmissions or in waiting times until the content is reliably delivered.

We capitalize on the coded packet transmission model for time variant channel proposed in [11] and depicted in Fig. 1 to characterize per-receiver completion time. This model is used to characterize per-receiver packet transmission adapted to its channel variation, and to characterize the network virtualized channel and the optimization of coded packets to be transmitted over it.

Therefore, according to [11], the end-to-end completion time from the GEO satellite source to a single receiver $k$ over a time variant channel is given as;

$$T(i, h_j) = T_d(N_i, h_j) + \sum_{l=1}^{i} P^{N_i}_{(i,h_j) \to (l,h_{j+N_i})} T(l, h_{j+N_i+1}), \quad (2)$$

with $T_d(N_i, h_j) = N_i T_p + T_w$, where $T_w$ is the waiting time for acknowledgment, $T_p$ is the packet time, and $N_i$ is the number of coded packets to be sent in batches for combining $i$ degrees of freedom or packets. The matrix $P$ is the one step transition matrix of the proposed model, where:

$$\left(\prod_{i=1}^{N_i} P\right)_{(i,h_j) \to (l,h_{j+N_i})} = P^{N_i}_{(i,h_j) \to (l,h_{j+N_i})},$$

corresponds to all the transition probabilities over the time slots from the initial state at $h_j$ until the state at $h_{j+N_i}$.

The index $j + N_i + 1$ appears in the delay term due to the acknowledgment.

In turn, a set of $K$ completion times will be associated to the receivers, suggesting a unifying approach that allow for a joint optimization of all the completion times of all the receivers. Given that such a problem is unsolvable, we propose two novel heuristic schemes that allow for near optimal joint optimization of the number of coded packets to multicast to all receivers.

### A. Maximum Erasure Scheme (MaxPe)

In this scheme, we propose to represent the receivers' channels in the multicast group by their joint global *MaxPe* encountered by each receiver over each time slot in the observation time window of channel variation. In turn, the optimization problem of such scheme can be written as:

$$\max_{P_e(h_k(t_1)),...,P_e(h_K(t_T))} \min_{N_1,...,N_i} T(i, h_j) = \\ \max_{P_e(h_k(t_1)),...,P_e(h_K(t_T))} \min_{N_1,...,N_i} T_d(N_i, h_j) + \\ \max_{P_e(h_k(t_1)),...,P_e(h_K(t_T))} \min_{N_1,...,N_i} \sum_{l=1}^{i} P^{N_i}_{(i,h_j) \to (l,h_{j+N_i})} T(l, h_{j+N_i+1}),$$
$$\forall k = \{1,..,K\} \in \tau = \{1,...,T\} \quad (3)$$

As it is known that the optimization problem is combinatorial and hard, we join the proposed *MaxPe* scheme to the heuristic Adaptive Network Coding (ANC) scheme in [11] to solve the problem of finding the number of coded packets to transmit to all receivers in the multicast group. Therefore, the set of coded packets that will be multicasted from the GEO satellite to all receivers is guaranteed to be decoded using such scheme with a design for worst receiver's channel condition. $N_1^*, ..., N_i^*$ is found by iterating over the vector of degrees of freedom as:

$$\sum_{s=j}^{N_i^*} (1 - P_e(h_s)) = i,$$
$$\forall P_e(s), \quad s.t. \ P_e(s) : \max\{P_e(h_k(t_1)), ..., P_e(h_K(t_T))\},$$
$$\forall k = \{1,..,K\} \in \tau = \{1,...,T\} \quad (4)$$

### B. Maximum Completion Time Scheme (MaxCT)

In this scheme, we propose to represent the receivers channels in the multicast group by the worst receiver's channel as a reference channel, which encounters *MaxCT* or maximum completion time to receive and decode reliably all coded packets from the GEO satellite within the observation time window of channel variation. In turn, the optimization problem of such scheme can be written as follows,

$$\max_k \min_{N_1,...,N_i} T(i, h_j) = \max_k \min_{N_1,...,N_i} T_d(N_i, h_j) \\ + \max_k \min_{N_1,...,N_i} \sum_{l=1}^{i} P^{N_i}_{(i,h_k) \to (l,h_{j+N_i})} T(l, h_{k+N_i+1}),$$
$$\forall k = \{1,..,K\} \quad (5)$$

As it is known that the optimization problem is combinatorial and hard, we resort to the heuristic ANC scheme in [11] to solve the problem of finding the number of coded packets to transmit to all receivers in the multicast group, therefore, the set of coded packets the GEO satellite will multicast with a guarantee that all receivers will be able to decode under the design for worst case condition $N_1^*, ..., N_i^*$ is found by iterating over the vector of degrees of freedom as follows,

$$\sum_{s=j}^{N_i^*} (1 - P_e(h_s)) = i,$$
$$\forall P_e(s), \quad s.t. \ P_e(s) : \{P_e(h_k(t_1)), ..., P_e(h_k(t_T))\},$$
$$k : \max CT_k, \quad \forall k = \{1,...,K\} \quad (6)$$

A similar approach of *maxCT* was proposed to address the XOR network coding multicast scenario for receivers encountering similar erasures in [16], however, the way the completion time is modeled and measured lacks the consideration of packet erasures dependency which is something present in the model in [11] with time variation. Therefore, besides the difference in the coding framework they used, such an assumption is absent in their work.

## IV. NUMERICAL RESULTS

In this section, we show numerical results obtained through computer simulations for evaluating the performance of the proposed schemes for coded multicast. The application scenario considers one GEO satellite and ten mobile receivers moving on the Earth ground. Thus, each link is characterized by a Round Trip Time (RTT) equals $0.2388$ s.

The receivers are randomly positioned so that each one experiences a different channel quality. To this aim we suppose that the receivers are moving in a random direction with a constant speed equals $5$ m/s.

In addition, the receivers are supposed to be located within a Low Height Building scenario [14], that considers the presence of three propagation states: line of sight, moderate and deep fading. To this aim, we also suppose that the receivers are equally distributed within the three states.

The performance is evaluated in terms of delay, throughput and average number of packets, by considering the on-board satellite transmitter multicasts a maximum batch of $i$ equals 10 data packets or degrees of freedom (dof), each of size $B$ equals $10^4$ bit. We consider data packets with a duration equal to $0.67$ ms.

The proposed coded multicast schemes are evaluated, and the results are compared with two benchmark schemes: the per-receiver ANC scheme, and per-receiver non-adaptive NC scheme, for time variant channels, proposed in [11].

*a) Maximum Erasure Scheme:* This scheme considers that we adapt to the worst channel conditions among all the receivers at any time instant. This corresponds to a virtual channel composed by the channel behavior having instantaneously the maximum erasure, and, hence, the higher attenuation. Such assumption can be seen as a virtual channel

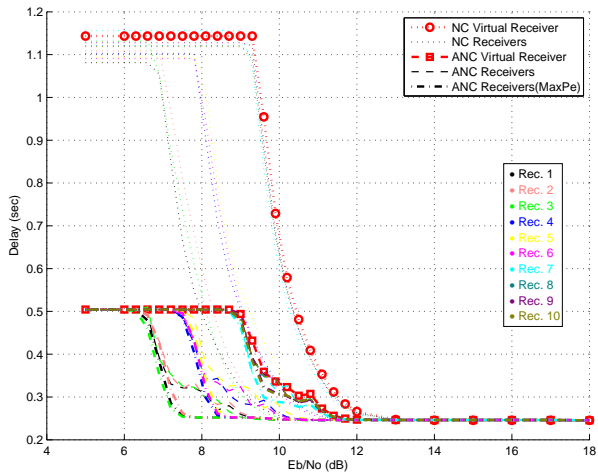

Fig. 2. Performance in terms of delay for the Maximum Erasure scheme by considering 10 receivers at different $E_b/N_0$ values.

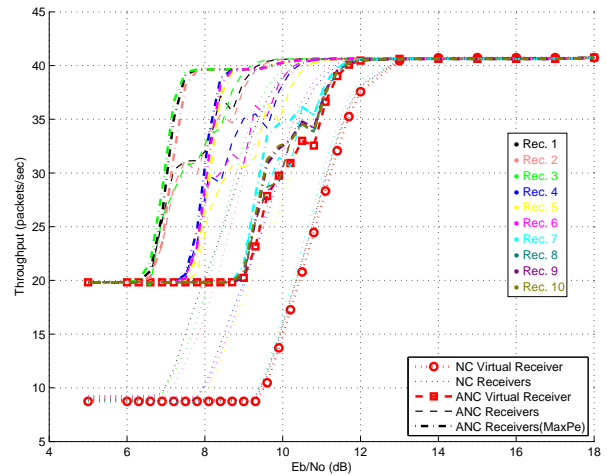

Fig. 3. Performance in terms of throughput for the Maximum Erasure scheme by considering 10 receivers at different $E_b/N_0$ values.

associated to the worst conditions among all in the multicast group.

Notice that in the figures, a certain color is used to express a certain receiver, and the same line type is used to express the scheme: dotted with/without circle represents the non-adaptive NC, dashed with/without rectangle represents the ANC, and the dotted-dashed represents the joint ANC with *MaxPe* scheme for coded multicast.

In Fig. 2 the delay performance is shown comparing the *MaxPe* policy adopting ANC, with the two benchmark ANC and NC schemes. Moreover, as a reference, the performance for the virtual receivers, employing NC, and ANC are also shown as a reference. It is possible to notice that the receivers are grouped into three main groups, reflecting the three different propagation conditions: line-of-sight, moderate shadowing, and deep fading. As expected, it is possible to see that NC encounters higher delays w.r.t. ANC for all receivers. Additionally, the receivers adopting the NC schemes selected by the Virtual receivers show a delay gain with respect to the performance obtained by using a pre-receiver policy. This can be observed by looking to the values in Tab. I, where, among other performance indicators, the average delay results for all the considered $Eb/N0$ values are reported for the 10 receivers.

Moreover, it is possible to notice that, due to their worst case condition design criterion, the *NC Virtual Receiver* and *ANC Virtual Receiver* experience the highest delay among all NC and ANC receivers.

Moreover, the following main observations are drawn:

- Receivers under line of sight and moderate fading, are gaining the most in terms of delay. For instance, Tab. I emphasizes the gain of receiver 5 equals $16.01$ ms.
- The receivers under deep fading, i.e., receivers 7 to 10, still enjoy gains compared to the case without the virtualization scheme albeit in a limited way.

In Fig. 3, the throughput performance is shown, in terms of delivered packets per second. Similar to the delay performance it is possible to notice that the receivers employing the *MaxPe* policy are gaining compared to the receivers employing a per-receiver ANC scheme. This has been highlighted in Tab. I where the average throughput for the 10 receivers is reported.

In general, the throughput performance of *MaxPe* receivers can show noticeable gains with respect to the per-receiver ANC scheme at the moderate SNR values.

Finally, the performance in terms of average number of packets is shown in Fig. 4. Those ave. no. of packets are drawn from evaluating the delay encountered at zero waiting time [12], when the system is adapted to the virtualized channel. Thus, the variance between the virtualization schemes is negligible, or small compared to no-virtualization. The gain in the delay and throughput performance at moderate SNR in cast to an increase in the average number of transmitted packets. Moreover, it is worth to observe that the cost associated to larger batches of coded packets, appears as a reward in the delay due to less retransmissions and less RTTs.

*b) Maximum Completion Time Scheme:* The performance evaluation of this virtualization scheme is applied considering the channel time-variant vector of the receiver suffering the maximum completion time among all receivers in the multicast group as the channel of a virtual receiver. Such virtual receiver becomes the reference receiver to all other receivers in the multicast group to which the coded transmission will be designed or adapted.

In Tab. I, it is possible to notice that, in the considered scenario, receiver 9 suffers the highest completion time. Hence, other receivers are supposed to utilize the ANC transmission strategy of receiver 9 and optimize their transmission of coded packets according to it. The performance of the *maxCT* scheme is compared with per-receiver NC and ANC benchmark schemes.

Fig. 5, illustrates the delay performance of the proposed *MaxCT* scheme and the benchmark schemes. Receiver 9 is the one used as the reference receiver for the multicast

TABLE I
SUMMARY TABLE OF THE PERFORMANCE FOR GEO SATELLITE RESULTS.

| Rec. No. | Channel gain [dB] | Delay [ms] | | | Throughput [packet/s] | | | Ave. No. of packets | | | Delay gain | | Throughput gain | |
|---|---|---|---|---|---|---|---|---|---|---|---|---|---|---|
| | | Without virtual. | MaxPe | MaxCT | Without virtual. | MaxPe | MaxCT | Without virtual. | MaxPe | MaxCT | MaxPe | MaxCT | MaxPe | MaxCT |
| 1 | 0,16 | 301,01 | 290,35 | 290,30 | 35,39 | 36,66 | 36,67 | 16 | 19 | 19 | 10,66 | 10,71 | 1,27 | 1,27 |
| 2 | 0,09 | 304,01 | 293,77 | 293,72 | 35,12 | 36,38 | 36,39 | 16 | 19 | 19 | 10,24 | 10,29 | 1,26 | 1,27 |
| 3 | 0,22 | 302,18 | 288,00 | 287,95 | 35,34 | 36,84 | 36,85 | 15 | 19 | 19 | 14,18 | 14,23 | 1,50 | 1,51 |
| 4 | -0,95 | 330,62 | 319,02 | 318,97 | 32,99 | 34,40 | 34,41 | 19 | 21 | 21 | 11,59 | 11,64 | 1,41 | 1,42 |
| 5 | -1,00 | 339,19 | 323,18 | 323,14 | 32,24 | 34,06 | 34,07 | 20 | 21 | 21 | 16,01 | 16,06 | 1,82 | 1,83 |
| 6 | -0,83 | 331,96 | 321,21 | 321,15 | 32,89 | 34,23 | 34,24 | 19 | 21 | 21 | 10,75 | 10,80 | 1,34 | 1,35 |
| 7 | -2,19 | 371,92 | 367,28 | 369,45 | 29,73 | 30,18 | 29,98 | 24 | 24 | 24 | 4,64 | 2,47 | 0,46 | 0,25 |
| 8 | -2,26 | 375,03 | 372,46 | 375,03 | 29,46 | 29,68 | 29,46 | 24 | 24 | 24 | 2,57 | 0,00 | 0,22 | 0,00 |
| 9 | -2,29 | 375,17 | 372,41 | 375,17 | 29,47 | 29,69 | 29,47 | 24 | 24 | 24 | 2,76 | 0,00 | 0,23 | 0,00 |
| 10 | -2,27 | 375,05 | 372,02 | 374,75 | 29,46 | 29,71 | 29,48 | 24 | 24 | 24 | 3,03 | 0,30 | 0,25 | 0,02 |

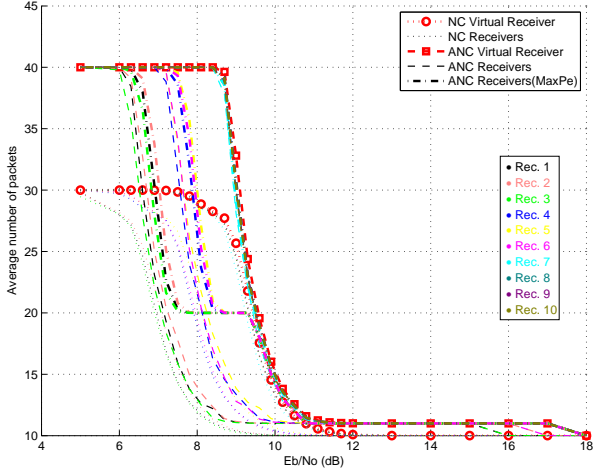

Fig. 4. Performance in terms of average number of packets for the Maximum Erasure scheme by considering 10 receivers at different $E_b/N_0$ values.

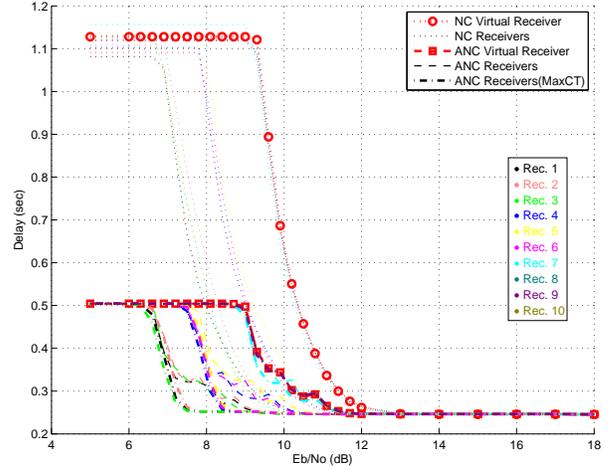

Fig. 5. Performance in terms of delay for the Maximum Completion Time scheme by considering 10 receivers at different $E_b/N_0$ values.

group. Similar to the previous set of results, the receivers are grouped into three main groups representing the three channel scenarios: line of sight, moderate fading, and deep fading.

Additionally, it is possible to see that using a receiver as a reference for setting up a multicast strategy allows to gain with respect to the per-receiver strategy. This has been clearly demonstrated as delay gains in Tab. I.

However, we observe for the receivers suffering deep fading (receivers 7-10), that *MaxPe* starts outperforming in delay/throughput gains the *MaxCT* scheme, and the gain collapse as the receiver has worst channel attenuation. This is due to the fact that the *MaxPe* leads to a design of larger batches of coded packets than *MaxCT*, however, in the rounded average to the nearest integer (in Tab. I), the two virtualization schemes have roughly similar coded packets.

Similarly, in Fig. 6, it is possible to see that by using the *MaxCT* scheme it is possible to gain with respect to the per-receiver ANC in terms of throughput. A similar trend is observed for receivers with deep fading with respect to the throughput.

Fig. 7 and Tab. I, show clearly an increase in the number of average coded packets for the *MaxCT* scheme. The more the receiver suffers deep fading, the more is such increase. However, as we discussed earlier, the *MaxCT* outperforms the *MaxPe* until the receivers encounter very deep fading where the later outperforms. In general, worth to observe that the cost in terms of resources comparing both virtualization schemes, is almost negligible.

Finally, Tab. II provides a summary of comparison of both virtualization schemes on their virtual channels. In particular, comparing non-adaptive schemes to adaptive ones using *MaxPe* and *MaxCT*, we can see a gain of $638.85$ ms and $623.48$ ms on the max mean completion time, respectively. This bear witness on the gains introduced from making a careful design of the adaptive coded packets, while guaranteeing through both virtualization schemes to have the receivers able to decode reliably.

We can see that the cost of adaptation paid is a maximum of 10 packets over non-adaptive ones for both virtualization schemes, a cost worth to be paid to serve almost half the maximum delay. This is again resorted that a transmission of longer adaptive batches is associated with less number of RTTs. 40 means that we need to transmit in average a batch of 40 coded packets to receive reliably 10. However,

TABLE II
PERFORMANCE TABLE OF THE PROPOSED VIRTUAL CHANNELS FOR GEO SATELLITE.

| Virtual channel scheme | Virtual receiver No. | Virtual receiver channel gain [dB] | | Non-adaptive Network Coding (NC) | | | | | | Adaptive Network Coding (ANC) | | | | | |
|---|---|---|---|---|---|---|---|---|---|---|---|---|---|---|---|
| | | | | Delay [ms] | | Thr. [packet/s] | | No. of packets | | Delay [ms] | | Thr. [packet/s] | | No. of packets | |
| | | Max. | Min. | Max. | Min. | Max. | Min. | Max. | Min. | Max. | Min. | Max. | Min. | Max. | Min. |
| MaxPe | [1,2,...,10] | -2,25 | -2,55 | 1143,25 | 245,50 | 40,73 | 8,75 | 30 | 10 | 504,4 | 245,5 | 40,73 | 19,83 | 40 | 10 |
| MaxCT | 9 | -2,15 | -2,55 | 1127,88 | 245,50 | 40,73 | 8,87 | 30 | 10 | 504,4 | 245,5 | 40,73 | 19,83 | 40 | 10 |

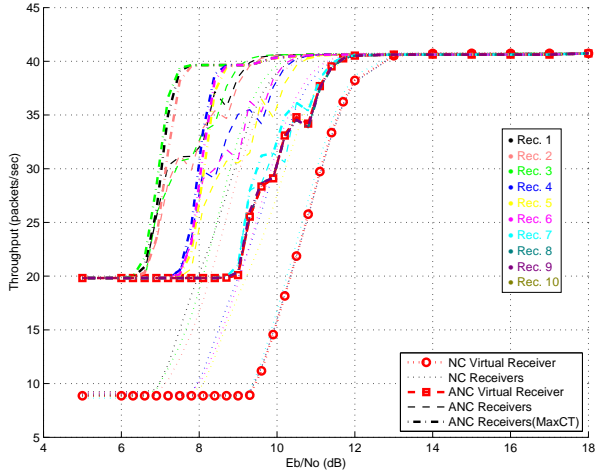

Fig. 6. Performance in terms of throughput for the Maximum Completion Time scheme by considering 10 receivers at different $E_b/N_0$ values.

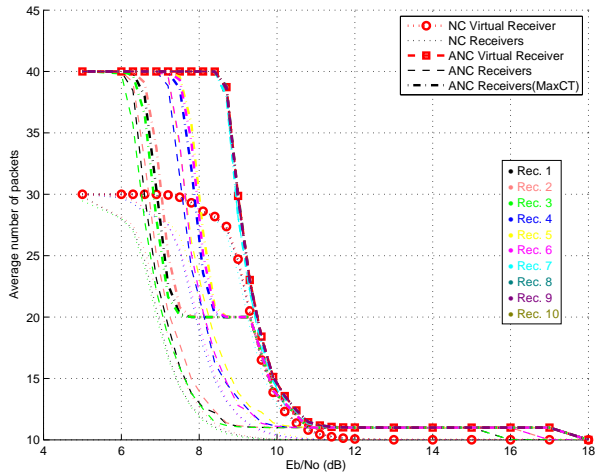

Fig. 7. Performance in terms of average number of packets for the Maximum Completion Time scheme by considering 10 receivers at different $E_b/N_0$ values.

30 average no. of coded packets for the non-adaptive scheme does not mean 10 data packets are guaranteed to be received, as there might not be sufficient degrees of freedom to decode all packets.

## V. CONCLUSIONS

We propose two network channel virtualization schemes to address the network coding for multicast with time variant channels. The proposed schemes rely on representing the multicast network with a worst performing virtual line network in packet erasure or completion time. The proposed virtualization schemes prove improvements when compared to per-receiver optimization, with and without adaptation, while assuring reliable reception by all receivers in the multicast group. Future research will consider correlation structure of the underlying multicast.